\documentclass{article}
\usepackage{mlspconf,amsmath,graphicx}
\usepackage{xcolor}
\usepackage{multirow}
\usepackage{ifsym}
\usepackage{enumitem}
\usepackage{hyperref}

\hypersetup{
    colorlinks=false,
    linkcolor=black,
    filecolor=black,      
    urlcolor=black,
}

\toappear{2024 IEEE International Workshop on Machine Learning for Signal Processing, Sept.\ 22--25, 2024, London, UK}

\title{TBDM-Net: Bidirectional Dense Networks with Gender Information for \\Speech Emotion Recognition }
\name{Vlad Strilețchi, Cosmin Strilețchi, Adriana Stan$^\dagger$\thanks{$^\dagger$Corresponding author: \texttt{adriana.stan@com.utcluj.ro}}}
\address{Communications Department,	Technical University of Cluj-Napoca, Romania\\
	   }

\begin{document}
%

\maketitle

\begin{abstract}

This paper presents a novel deep neural network-based architecture tailored for Speech Emotion Recognition (SER). The architecture capitalises on dense interconnections among multiple layers of bidirectional dilated convolutions. A linear kernel dynamically fuses the outputs of these layers to yield the final emotion class prediction. This innovative architecture is denoted as \textbf{TBDM-Net}: \textbf{T}emporally-Aware \textbf{B}i-directional \textbf{D}ense \textbf{M}ulti-Scale \textbf{Net}work. We conduct a comprehensive performance evaluation of \texttt{TBDM-Net}, including an ablation study, across six widely-acknowledged SER datasets for unimodal speech emotion recognition. Additionally, we explore the influence of gender-informed emotion prediction by appending either golden or predicted gender labels to the architecture's inputs or predictions. The implementation of \texttt{TBDM-Net} is accessible at: \url{https://github.com/adrianastan/tbdm-net}.
\end{abstract}
\begin{keywords}
speech emotion recognition, dense nets, convolutions, bidirectional layers, SER
\end{keywords}

\section{Introduction}
\label{sec:intro}

\begin{figure*}[th!]
\centering
\includegraphics[width=0.98\linewidth]{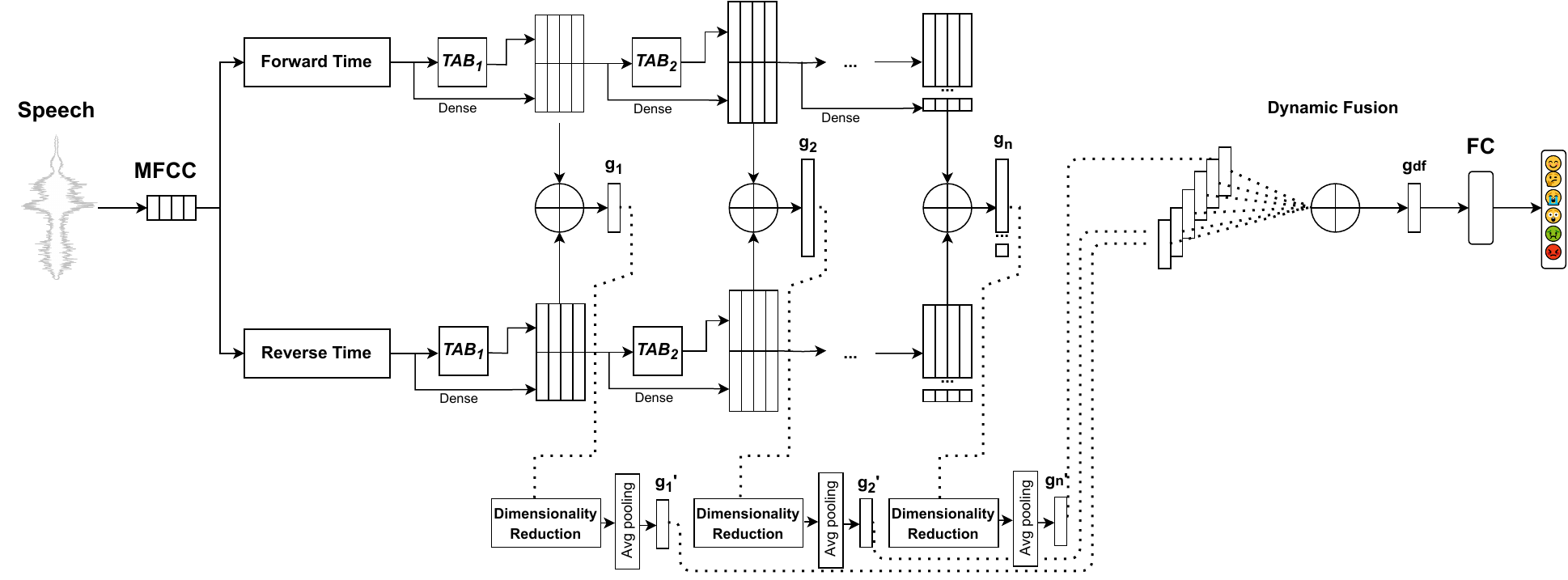}
\vspace{-0.4cm}
\caption{The TBDM-Net architecture. The forward and reverse time speech representations are passed through a series of Temporally-Aware Blocks (TABs). The intermediate bidirectional representations are concatenated ($g_k$), passed through a dimension reduction convolutional block and averaged to obtain a final concatenation of different time-scale representations ($g_k'$). Each TABs input is passed forward to the next modules through dense concatenative connections. The final representation dynamically fuses the multi-scale representations ($g_{df}$), and is passed through a fully connected layer ($FC$) to output emotion class probabilities.}
\label{fig:arch}
\end{figure*}

\begin{table*}[h!]
\centering
\footnotesize
\begin{tabular}{|llccccc|}
\hline
\textbf{Dataset} & \textbf{Emotions} & \textbf{Language} & \textbf{No. samples} & \textbf{Duration}  & \textbf{Speakers}  & \textbf{M/F}\\ \hline
\textbf{CASIA}~\cite{casia} & angry, fearful, happy, neutral, sad, surprised & Chinese & 1200 & 43' & 8 & 4/4\\ \hline
\textbf{EMOVO}~\cite{emovo} & angry, disgusted, fearful, happy, neutral, sad, surprised & Italian& 588 & 30' & 6 & 3/3\\ \hline
\textbf{EMODB}~\cite{emodb} & angry, bored, disgusted, fearful, happy, neutral, sad & German & 535 & 24' & 10 & 5/5\\ \hline
\textbf{IEMOCAP}~\cite{iemocap} & angry, happy, neutral, sad & English &5531 & 11h 37' & 10 & 5/5\\ \hline
\textbf{RAVDESS}~\cite{ravdess} & angry, calm, disgusted, happy,   fearful, sad, surprised & English & 1440 & 1h 28' & 24 & 12/12\\ \hline
\textbf{SAVEE}~\cite{savee} & angry, disgusted, fearful, happy, neutral, sad, surprised & English &480 & 30' & 4 & 4/0\\
\hline
\end{tabular}
\caption{Evaluated SER datasets. The M/F column reports the number of male and female speakers in the dataset.}
\label{tbl:datasets}
\end{table*}

Speech emotion recognition (SER) plays an important role in adding a new dimension to speech-enabled applications. As opposed to the linguistic content which can be to a large extent controlled by the speaker, emotions pose a harder task when the speakers intends to  limiting or control them in conversations. 
Understanding emotions expressed through speech can significantly improve human-computer interaction, enabling machines to respond more appropriately to user needs and preferences.
However, a major problem in SER-related tasks is still the lack of accurate training data, as many of the available datasets contain only acted speech, i.e. various emotions are elicited by actors in a controlled scenario. Thus, benchmarking\footnote{\url{https://emosuperb.github.io/}} SER can still pose challenges for real-life data deployments and evaluation. 

As is the case for many other research fields, SER systems are to a large extent based on deep neural networks with increasingly complex architectures. An important research direction is also the use multimodal (e.g. text, image, audio or video) characteristics. However, for this study we rely solely on the information available within the speech signal. 

Some of the most notable results in this area are those reported by~\cite{lightsernet}. Light-SERNet focuses on reducing the computational complexity of SER and uses a network based only on convolutional layers and local feature learning blocks.
Convolutional networks were also used by~\cite{7472669,7820699}.
Wen et al.~\cite{cpac} introduce an architecture based on capsule nets and transfer learning. It also deals with cross-corpus evaluation, for which an adversarial module is used. Gradient-based adversary learning framework that jointly estimates the SER labels while also normalising speaker characteristics is proposed in \cite{9747460}. The method is based on large pretrained models and also analyses the use of small labelled datasets.
Croitoru et al.~\cite{croitoru2022lerac} describe a novel general method for learning rate adjustments during the training process where the learning rate of the layers closer to the network's output are adjusted more finely. The proposed method is evaluated over several distinct prediction tasks, including speech emotion recognition. The architecture used in SER is based on a transformer and a dense net structure. 

Schuller et al.\cite{shuler} attempt to preserve the emotional saturation of speech frames by employing frame-level speech features and attention-based LSTM recurrent neural networks. Recurrent layers were also adopted by \cite{10128692}, and combined with Wavelet transforms and 1D CNNs to extract multiresolution representations of the input speech. 
The Dual-TBNet model of \cite{dual-tbnet} proposed the combination of 1D convolutional layers, transformer, and BiLSTM modules, to maximise the robustness of the independent speech abstractions which are then fused to provide the final emotion decision. 
Notably, Ye et al.~\cite{ye2023temporal} introduce \texttt{TIM-Net}, which outperforms previous methods on six standard SER datasets. Inspired by this, we draw from their work and implement several modifications to enhance overall emotion prediction accuracy.


Building upon prior research, this paper introduces a novel architecture for Speech Emotion Recognition (SER) classification. The architecture employs temporally-aware bidirectional dense networks, referred to as Temporally-Aware Bi-directional Dense Multi-Scale Network (\textbf{\texttt{TBDM-Net}}). The primary contributions of the paper can be summarised as follows: (i) the introduction of a new deep architecture for SER; (ii) an assessment of the proposed architecture across six multilingual SER datasets; (iii) an ablation study to analyse the impact of each architectural module on final performance; and (iv) an examination of the influence of speaker gender information on emotion classification accuracy.

\section{TBDM-Net Architecture}
\label{sec:net}

\texttt{TBDM-Net}
uses a series of temporally-aware convolution blocks (\texttt{TABs}) with incremental dilation coefficients over the direct and inverse time representation of the input speech signal. All forward and reverse temporal blocks are densely connected. These connections enable the network to exploit previously computed features and reduce the number of required parameters~\cite{huang2018densely}.
The intermediate representations obtained from each TAB are summed with their reverse time correspondents.  The concatenation of the forward and reverse time temporal blocks' output is passed through a dimension reduction convolutional layer, such that all intermediate representations have the same dimension. 
All TAB representations are then concatenated and dynamically fused. The final emotion prediction probabilities are obtained through a simple feed forward layer.

The activation function employed within the Temporal Attention Blocks (TABs) is the Gaussian Error Linear Unit (GELU)~\cite{hendrycks2020gaussian}. GELU is characterised by its smoothness, maintaining a continuous first derivative across its range, which fosters stable and efficient optimisation processes. Its attribute of granting non-zero gradients for both positive and negative inputs facilitates unrestricted information flow during both forward and backward propagation. Additionally, GELU has demonstrated efficacy in addressing the vanishing gradient issue in deeper neural networks.

The complete architecture is presented in Figure~\ref{fig:arch}. Its implementation is available at: \url{https://github.com/adrianastan/tbdm-net}.



\section{Evaluation}
\label{sec:results}

\begin{table*}[th!]
\centering
\footnotesize
\begin{tabular}{|llccc||llccc|}
\hline
\textbf{Dataset} & \textbf{Model} & \textbf{UAR}$\uparrow$ & \textbf{WAR}$\uparrow$ & \textbf{F1}$\uparrow$ & \textbf{Dataset} & \textbf{Model} & \textbf{UAR}$\uparrow$ & \textbf{WAR}$\uparrow$ & \textbf{F1}$\uparrow$\\ \hline \hline
\multirow{4}{*}{\textbf{CASIA}}& Dual-TBNet~\cite{dual-tbnet} & \textbf{95.70} & - & \textbf{95.80} & \multirow{4}{*}{\textbf{EMODB}} & Dual-TBNet~\cite{dual-tbnet} & 84.10 & - & 84.30 \\ 
    & TIM-Net~\cite{ye2023temporal} & 89.60 & 90.40 & 89.90 & 
    & TIM-Net~\cite{ye2023temporal}& 91.09 & 89.30 & 89.00 \\ 
                                 & \textbf{TBDM-Net::BT}  & 86.54 & 85.66 &  85.77   &                                   & \textbf{TBDM-Net::BT} &  90.01 & 88.23 & 88.30\\
                                 & \textbf{TBDM-Net::300} & 91.01 & \textbf{90.50}  & 90.54 &                                 & \textbf{TBDM-Net::300} & \textbf{92.94} & \textbf{91.40} & \textbf{91.55}\\
\hline \hline

\multirow{4}{*}{\textbf{EMOVO}} & 1BTPDN~\cite{9226478} & 74.31 & - & - & \multirow{4}{*}{\textbf{IEMOCAP}} & SCFA~\cite{zhao23e_interspeech} & 67.91 & - &  66.42 \\ 

  & TIM-Net~\cite{ye2023temporal}& 84.60 & 85.80  & 84.10 & 
  & TIM-Net~\cite{ye2023temporal}& 68.70 & 67.90 & 67.40 \\
                                 &  \textbf{TBDM-Net::BT}& 84.20 & 82.12 &  82.05 &                             & \textbf{TBDM-Net::BT} & 71.78 & 70.05 & 70.19\\
                                 & \textbf{TBDM-Net::300} & \textbf{88.10} &  \textbf{87.06} & \textbf{87.19}  &                                &\textbf{TBDM-Net::300} & \textbf{73.28} & \textbf{71.88} & \textbf{71.94}\\
\hline \hline

\multirow{4}{*}{\textbf{RAVDESS}} & WMA~\cite{10128692}&  81.40 & 81.20 & - & \multirow{4}{*}{\textbf{SAVEE}} & Dual-TBNet~\cite{dual-tbnet} & \textbf{83.30} &  - &  \textbf{82.10} \\

                                & TIM-Net~\cite{ye2023temporal} & 88.00 & 89.30 & 88.10 & 
                                &  TIM-Net~\cite{ye2023temporal}& 80.09 & \textbf{81.19} & 78.20 \\
                                 & \textbf{TBDM-Net::BT} & 85.29 & 84.30 &  84.40  &                          & \textbf{TBDM-Net::BT} & 78.47 & 77.49 & 77.90\\ 
                                 & \textbf{TBDM-Net::300} & \textbf{91.60} & \textbf{90.97} & \textbf{91.02}  &&\textbf{TBDM-Net::300} & 81.21 & 80.41 & 80.60\\
\hline
\end{tabular}
\caption{TBDM-Net's evaluation. The results are based on the average performance over a 10-fold cross validation report on the unweighted average recall (\textbf{UAR}), weighted average recall (\textbf{WAR}), and weighted \textbf{F1}-score. \texttt{TBDM-Net-BT} is the best model in terms of training set $WAR$, while \texttt{TBDM-Net-300} is the model obtained at the end of the 300 epoch training process. Best results are marked in boldface. \vspace{0.1cm}}
\label{tbl:results}
\end{table*}

\subsection{Speech datasets and features}

For the evaluation
we used six standard speech emotion recognition datasets 
An overview of these datasets is shown in Table~\ref{tbl:datasets}. We note that for the IEMOCAP dataset, due to the high class imbalance in the original dataset, we select only 4 classes, and relabel the \emph{excited} class into \emph{happy}.\footnote{Common approach for the IEMOCAP dataset in other studies.} The final subset contains 5531 speech samples.

From Table~\ref{tbl:datasets} it can be noticed that there is no complete overlap between all sets of emotions annotated or rendered within the speech corpora. There is also a high variation among the amount of speech data, as well as the number and gender of the speakers. 
One other important characteristic of this dataset selection is its multilingual aspect: IEMOCAP, RAVDESS and SAVEE are English datasets, CASIA is Chinese, EMOVO is Italian, and EMODB is German. All these features of the data selection enable us to perform a thorough analysis of the proposed architecture across several dimensions of speech and emotion variation. 

Similar to~\cite{ye2023temporal}, we use 39 Mel frequency cepstral coefficient (MFCC) representations extracted with the default settings in the Librosa module.\footnote{\url{https://librosa.org/}} A fixed number of frames, different across the datasets, are used to represent the utterances. 
Shorter utterances are zero-padded left and right, while longer utterances are cropped to a central segment.\footnote{Feature sets are available from the authors of~\cite{ye2023temporal}: \url{https://github.com/Jiaxin-Ye/TIM-Net_SER}} The maximum temporal dimension of the MFCC representations is between 172 frames for the CASIA dataset, and 606 frames for the IEMOCAP dataset.


\subsection{Objective measures and training procedure}




Because of the varying sample sizes and emotions within each dataset, we opted to assess the performance of our network independently on each speech dataset through a 10-fold cross-validation process. The folds are chosen randomly, without any dependency on speaker or emotion.

As objective measures, we used the \textbf{unweighted average recall ($UAR$)},  \textbf{weighted average recall ($WAR$)} and \textbf{F1-score} ($F1$).
In standard implementations of multi-class classification, the $UAR$ is equal to the global prediction accuracy.\footnote{\url{https://www.evidentlyai.com/classification-metrics/multi-class-metrics}} The $WAR$ is a weighted measure which takes into account the number of samples from each class within the test set. $UAR$, $WAR$ and $F1-score$ enable us to directly compare our results against the previously published methods. 

The \texttt{TBDM-Net}'s architecture uses 6 temporal blocks for all speech datasets.
The dilation rates are: 1, 2, 4, 8, 16, and 32, respectively. The number of filters for each convolution is equal to the number of MFCC coefficients (i.e. 39), and use a kernel size of 2. 
The models were trained for 300 epochs, using an ADAM optimiser with a learning rate of $1e-3$, and $betas=(0.93, 0.98)$. The batch size was set to 64. No early stopping or learning rate scheduler were used. 
The best model across the 300 epochs in terms of training set $WAR$ was saved, and used to provide the intermediate evaluation results for the respective fold. A separate result is extracted from the model obtained after 300 epochs of training.\footnote{No major accuracy improvements were obtained beyond the 300 epoch training step.}

\subsection{Baseline results}

The results of \texttt{TBDM-Net}'s evaluation are shown in Table~\ref{tbl:results}. The table includes the performance of \texttt{TBDM-Net} as evaluated in two scenarios: (i) the best model in terms of training subset \texttt{WAR} (\texttt{TBDM-Net::BT}); and (ii) the model obtained after 300 epochs of training (\texttt{TBDM-Net::300}). The table also introduces the recomputed results for the \texttt{TIM-Net} architecture using the authors' official implementation.\footnote{These results differ from the ones introduced in the original paper, as the author's initial evaluation made use of the test set data for the best model selection.} The same set of input features and number of training epochs as for \texttt{TBDM-Net} were used.
We also present the top performance measures for each dataset as reported by previously published peer-reviewed methods. 

Observations reveal that \texttt{TBDM-Net} demonstrates heightened performance across nearly all assessed datasets. We posit that facilitating connectivity between the Temporal Attention Blocks (TABs) at varying resolutions and incorporating bidirectional paths enhances the capture of emotional abstraction. However, exceptions arise with the CASIA and SAVEE datasets, where the Dual-TBNet architecture distinctly outperforms. Further analysis is necessary to ascertain the reasons behind the diminished accuracy of \texttt{TBDM-Net} on these datasets.
A notable increase in performance is obtained for the EMOVO, RAVDESS and IEMOCAP datasets in terms of $UAR$--with over 3\% absolute recall increase. It is important to notice that IEMOCAP is the largest dataset, and the one which poses most challenges across all published studies in SER. These results are encouraging, as we are planning to use the proposed architecture in real-life scenarios where large volumes of data and less well-defined emotion elicitation would be found. 

\begin{table}[t!]
\footnotesize
\centering
\begin{tabular}{|lcccc|c|} 
\hline
    & \textbf{ReLU} & \textbf{w/o BD} & \textbf{w/o MS} & \textbf{5 TABs} & \textbf{TBDM-Net}\\
    \hline
    \textbf{UAR} & 90.76 & 90.09 & 90.81& 91.08& \textbf{91.60}  \\
    \textbf{WAR} & 90.34 & 89.99 & 90.69& 90.90& \textbf{90.97}\\
    \textbf{F1} & 90.42 & 90.06  & 90.80& 90.84& \textbf{91.02}\\
\hline
\end{tabular}
\caption{Ablation study on the \textbf{RAVDESS} dataset. The results are based on a 10-fold cross validation of the model trained for 300 epochs. Best results are marked in boldface.}\vspace{0.2cm}
\label{tbl:ablation}
\end{table}

\vspace{-.2cm}
\subsection{Ablation study}
We also introduce the results of an ablation study over the RAVDESS dataset. Similar results were found for the other datasets, as well. The study includes the following modifications made to \texttt{TBDM-Net}, with all the other hyperparameters of the architecture and training step having been frozen:

\begin{itemize}[leftmargin=*]
\itemsep-0.2em 
\item{{activation function} (\texttt{ReLU})} - we use the ReLU activation function in the TABs instead of GELU;
\item{{directionality} (\texttt{w/o BD})} - we compare the bidirectional(BD) network with a forward time-only network variation;
\item{{multi-scale} (\texttt{w/o MS})} - our proposed model uses multi-scale (MS) fusion of the intermediary states between TABs. We compare this with a variation which only uses the last state, disregarding the others;
\item{{number of TABs} (\texttt{5 TABs})} - the proposed model uses 6 TABs, and we compare it with a variation containing only 5 TABs.
\end{itemize}

The results are shown in Table~\ref{tbl:ablation}. 
It can be noticed that in terms of both $UAR$ and $WAR$, all modifications to the network yield performance increments, with the number of TABs having the least impact over the final results. This is encouraging especially if the network needs to be optimised for real-time applications. 
Most improvement is obtained from the use of bidirectional modules.

\subsection{Gender-informed results}

Given that previous literature strongly motivates the use of gender-differentiated systems in SER-related tasks~\cite{6853745,9222412}, we perform a similar evaluation for the \texttt{TBDM-Net} architecture. We also examine how real-life applications would perform the task when the speaker gender is also estimated from the input speech. 
Therefore, we first build a gender classifier based on large pretrained models' derived speech embeddings. Previous studies~\cite{math10213927} over such embeddings showed that the TitaNet architecture~\cite{titanet} exhibits a high correlation between its embeddings and the speaker's gender. The TitaNet-L\footnote{\url{https://catalog.ngc.nvidia.com/orgs/nvidia/teams/nemo/models/titanet_large}} pretrained model was selected and the corresponding embeddings extracted. To ensure a good classification performance, several classifiers were trained on the development partition of the VOXCELEB dataset~\cite{Nagrani_2017}, and tested on the VOXCELEB test partition, as well as the RAVDESS dataset. The results are shown in Table~\ref{tbl:clasif}. The SVC-based gender classifier's prediction in binary or probabilistic formats was used in the follow-up evaluation.

\begin{table}[bh!]
\centering
\footnotesize
\begin{tabular}{|l|c|c|}
    \hline
    & \multicolumn{2}{c|}{\textbf{Accuracy [\%]} $\uparrow$}  \\
    \textbf{Classifier} & \textbf{VOXCELEB}  & \textbf{RAVDESS} \\
    \hline
    Logistic Regression & 93.45 & 81.56 \\
    MLP Classifier & 99.83 & 82.70 \\
    Random Forest& 98.85 & 87.32 \\
    XGBoost & 98.15  & 90.90\\   
    Support Vector Classifier & \textbf{99.97} & \textbf{98.26}\\ \hline
\end{tabular}
\caption{Gender classifier accuracy evaluation. The classifiers were trained on the \emph{VOXCELEB-dev} dataset and evaluated on the \emph{VOXCELEB-test} and \emph{RAVDESS} datasets. Best results are marked in boldface.}
\label{tbl:clasif}
\end{table}

\begin{table}[bh!]
\centering
\footnotesize
\begin{tabular}{|l|c|c|c|}\hline 
    \textbf{System} & \textbf{UAR}$\uparrow$ & \textbf{WAR}$\uparrow$ & \textbf{F1}$\uparrow$ \\
    \hline\hline
    \texttt{baseline:full} & 91.60 & 90.97 & 91.02 \\ \hline
    \texttt{baseline:M} & 87.02 & 85.00 & 84.91 \\
    \texttt{baseline:F} & 91.23 & 90.13 & 90.31 \\
    \hline \hline
    \texttt{split:M} & 90.86 & 89.44 & 89.24 \\
    \texttt{split:F} & 90.71 & 90.41 & 90.71\\
    \hline \hline
    \texttt{post-hoc:golden-labels} & 90.75 & 90.41 & 90.44\\
    \texttt{post-hoc:binary} & 89.83 & 89.37 & 89.38\\
    \texttt{post-hoc:probabilities} & 90.04 & 89.51 & 89.50 \\
    \hline \hline
    \texttt{pre-hoc:golden-labels} & 91.34 & 91.18 & 91.22 \\
    \texttt{pre-hoc:binary} & 91.07 & 90.90 & 90.88 \\
    \texttt{pre-hoc:probabilities} & \textbf{91.70} & \textbf{91.31} & \textbf{91.33}\\    \hline 
\end{tabular}
\caption{Gender-informed TBDM-Net 10-fold cross-validation results on the \textbf{RAVDESS} corpus. Best results are marked in boldface.}
\label{tbl:ravdegen}
\end{table}

\begin{table}[th!]
\centering
\footnotesize
\begin{tabular}{|l|c|c|c|} \hline
    \textbf{System} & \textbf{UAR}$\uparrow$ & \textbf{WAR}$\uparrow$ & \textbf{F1}$\uparrow$ \\
    \hline \hline
    \texttt{baseline:full} & 73.28 & 71.88 & 71.94\\ \hline
    \texttt{baseline:M} & 72.72 & 70.53 & 70.75 \\
    \texttt{baseline:F} & 73.73 & 72.61 & 72.59 \\
    \hline \hline
    \texttt{split:M} & 72.62 & 70.05 & 70.23 \\
    \texttt{split:F} & 72.58 & 71.40 & 71.65 \\
    \hline \hline
    \texttt{post-hoc:golden-labels} & 72.03 & 70.72 & 70.88 \\
    \texttt{post-hoc:binary}& 71.45 & 70.18 & 70.33 \\
    \texttt{post-hoc:probabilities} & 71.69 & 70.43 & 70.60 \\
    \hline \hline
    \texttt{pre-hoc:golden-labels} & 73.68 & 71.94 & 71.97 \\
    \texttt{pre-hoc:binary} & 73.47 & 71.77 & 71.84 \\
    \texttt{pre-hoc:probabilities} & \textbf{73.69} & \textbf{72.22} & \textbf{72.26} \\\hline
\end{tabular}
\caption{Gender-informed TBDM-Net 10-fold cross validation results on the \textbf{IEMOCAP} corpus. Best results are marked in boldface.}
\label{tbl:iemogen}
\end{table}

The speaker's gender information was added to the \texttt{TBDM-Net} architecture either as post-hoc boosting information or concatenated to the input MFCC representations over the coefficients' dimensions. The results are reported over the RAVDESS and IEMOCAP datasets. 11 SER systems are evaluated in terms of 10-fold cross-validation, and their results are shown in Tables~\ref{tbl:ravdegen}~and~\ref{tbl:iemogen}. They pertain to the following setups:

\begin{itemize}[leftmargin=5.5mm]
\setlength\itemsep{-0.5em}
\item \texttt{baseline:full} - model trained on the entire train partition and evaluated on the entire test partition; 
\item \texttt{baseline:M} - model trained on the entire train partition and evaluated on the male subset of the test partition; 
\item \texttt{baseline:F} - model trained on the entire train partition and evaluated on the female subset of the test partition; 
\item \texttt{split:M} - the male subset of the dataset split into train and test partitions. 
\item \texttt{split:F} - the female subset of the dataset split into train and test partitions. 
\item \texttt{post-hoc:golden-labels} - based on the golden label for the speaker's gender, the corresponding gender-dependent model is used to make the final emotion prediction;
\item \texttt{post-hoc:binary} - the binary prediction of the pre-trained gender classifier is used to select the appropriate gender-dependent model;
\item \texttt{post-hoc:probabilities} - the probabilistic predictions of the gender classifier are used to weight the predictions of the two gender-dependent models;
\item \texttt{pre-hoc:golden-labels} - the golden label for the gender is concatenated to the MFCC representation at input;
\item \texttt{pre-hoc:binary} - the binary prediction of the gender classifier is concatenated to the MFCC representation at input;
\item \texttt{pre-hoc:probabilities} - the probabilistic output of the gender classifier is concatenated to the MFCC representations at input.
\end{itemize}

It is apparent that incorporating gender information into the input features as gender probability vectors leads to enhanced performance on both evaluated datasets. Nevertheless, the enhancements are only incremental, suggesting the need for a more thorough analysis and a refined feature merging strategy.

\section{Conclusions}
\label{sec:conclusions}

In this paper we introduced a novel densely connected deep architecture with temporally-aware blocks, which is able to perform accurate unimodal speech emotion recognition. The experimental results show that \texttt{TBDM-Net} outperforms the state-of-the art results on 6 multilingual SER datasets. An ablation study revealed that the most important aspect of the architecture is the use of bidirectional representations of the spoken output. However, this bidirectionality incurs additional complexity when used in real-time applications. In a separate evaluation, we investigated the effects of gender information over the model's performance. Results showed that \texttt{TBDM-Net} is to a large extent invariant to the speaker's gender, and that adding such information to the network only slightly improves the overall accuracy of the SER system.

In our future efforts, we aim to utilise the existing architecture for call-centre interactions by consolidating certain emotion categories found in the datasets. Additionally, we seek to explore whether further adjustments to the network could streamline its computational demands, making it more suitable for real-time usage. Crucially, we plan to assess the algorithms' performance across different datasets to evaluate their predictive capabilities beyond their original distribution.




\bibliographystyle{IEEEbib}
\bibliography{main}

\end{document}